\documentclass[11pt,a4paper]{article}

\usepackage[cp1251]{inputenc}
\usepackage[T2A]{fontenc}

\usepackage[pdftex]{graphicx}
\usepackage{hyperref}
\usepackage{amsmath}
\usepackage{amssymb}
\usepackage{amsxtra}
\usepackage{amsthm}
\usepackage{latexsym}
\usepackage{array}
\usepackage{multirow}
\usepackage{hhline}
\usepackage[in]{fullpage}
\usepackage{indentfirst}

\def\sgn{\mathrm{sgn}\,}

\usepackage{hyperref}
\theoremstyle{plain}
\newtheorem{theorem}{Theorem}
\begin{document}

\begin{center}
\begin{Large}
\textbf{Dynamics of a few vortices in an ideal fluid and in a Bose-Einstein condensate}
\vskip 5mm
\textbf{Pavel\,\,E.~Ryabov$^{1,2,3}$, Sergei\,\,V.~Sokolov$^{2,3}$ and Gleb\,\,P.~Palshin$^{1}$}
\end{Large}
\\[5mm]

${}^{1}$\! Financial University under the Government of the Russian Federation\\
Leningradsky prosp. 49, Moscow, 125993, Russian Federation\\

${}^{2}$\! Moscow Institute of Physics and Technology (National Research University)\\
Institutskiy per. 9, Dolgoprudny, Moscow Region, 141701, Russian Federation\\

${}^{3}$\! Mechanical Engineering Research Institute, Russian Academy of Sciences\\
Maly Kharitonyevsky per. 4, Moscow, 101990, Russian Federation\\[5mm]

E-mail: PERyabov@fa.ru, sokolov.sv@phystech.edu, gleb.palshin@yandex.ru

\end{center}

\begin{abstract}
Completely Liouville integrable Hamiltonian system with two deg\-rees of freedom is considered. This Hamiltonian system describes the dynamics of two vortex filaments in a Bose\,-- Einstein condensate enclosed in a cylindrical trap and dynamics of two vortices in ideal fluid contained in a domain with the form of  circular cylinder, generatrix of which  is parallel to vortices. An explicit reduction to a system with one degree of freedom for the case of arbitrary signs of the vortex intensities is derived from the original system under consideration. Along with the fact that in the case of two vortices with intensities of opposite signs, a bifurcation diagram was obtained for a generalized two-parameter integrable family, similar to that obtained earlier for the case of a Bose-Einstein condensate,
a new bifurcation diagram was found that had not been encountered before. Various types of vortex dynamics corresponding to points inside chambers of bifurcation diagrams are presented.
\end{abstract}

\noindent\textit{Keywords}:\,
{vortex dynamics, Bose\, -- Einstein condensate, completely integrable Ha\-mil\-to\-nian systems, bifurcation diagram of momentum map, bifurcations of Liouville tori}\\

\textit{MSC2010 numbers}: 76M23, 37J35,  37J05, 34A05 

\textit{Received March 22, 2021}

\section{Introduction}
\setcounter{equation}{0}

In this paper, we consider one of the integrable problems of vortex dynamics. The first integrable cases in the dynamics of vortices were found almost immediately after the discovery of the theory of vortices by Helmholtz \cite{helm}. For example, the classical three-vortex problem was investigated by Kirchhoff \cite{kirh} and Gr{\"{o}bli \cite{greb}, a little later Greenhill indicated the integrability of the two-vortex problem near the cylinder \cite{greenh}, and numerous later results of general and special cases can be mentioned integrability described in monograph of Goryachev \cite{goryach} in the classical period. In modern works, these classic integrable cases have been redefined from different perspectives. An overview, for example, of the results on the three vortex problem can be found in the work of Aref \cite{aref}, where, by the way, this problem was rediscovered.
Numerous literature is devoted to this problem, but it is covered in more detail in the book of Borisov, Mamaev \cite{bormamMatMet}, which, unfortunately, is available only in Russian. In the modern period, it should be noted the work of three vortices in the sphere: Borisov, Pavlova \cite{BorPav}, Borisov, Lebedev \cite{BorLeb}, Newton \cite{newt}, the theory of vortex chains on the sphere, described in Borisov, Mamaev \cite{bormamVortChain}, which continues Goryachev’s research, as well as some similar problems on the interaction of vortices on the torus and on the cylinder, as outlined by Stremler and Aref \cite{StremAref}.

From a hydromechanical point of view, these problems are usually considered at the general physics manner. Modern methods for topological analysis of these problems were first identified in the works of Borisov, Pavlov, Lebedev, in which three vortices were considered on a plane and on a sphere, bifurcation diagrams were constructed, and various special cases were analyzed. It is curious that, along with studies of bifurcations in the three vortex problem, it is of interest to study the topology of the symplectic leaf given in Borisov, Mamaev, Bizyaev \cite{bormambiz}. Note that the analysis performed in the works of Borisov, Pavlov, Lebedev from a topological point of view is not exhaustive. During this time, new results appeared on the connection between the bifurcation complex and the stability of critical movements, for example, \cite{BolBorMam1}. The method is proposed for the study of stability and the role of bifurcation diagrams is recognized for a general understanding of bifurcations of the dynamic behavior, stability analysis and many other facts, even analysis of absolute movement, including choreography. Recently, a series of works on three vortices has appeared, one of which is fixed \cite{RyzhKosh}.

In this paper, we consider an unusual analogue of the Greenhill problem of two vortices of arbitrary intensity, of opposite signs, moving around a cylinder. This analogue differs significantly from the classical setting in that a new free term appears, which simplifies the consideration, since it is possible to obtain the discriminant curve in an explicit form. For completeness, we also consider the general case of a two-parameter family of integrable systems, covering both the classical ideal fluid and the case of the Bose-Einstein condensate, in order to compare the dynamics of these two problems and the bifurcations of the tori that arise on this path. In a sense, the classical problem is more complicated than the corresponding problem for condensate.

The mainstream of the vortex analytical dynamics is an integrable models of point vortices on a plane. Studies of the dynamics of vortices in a quantum liquids, have shown that quantum vortices behave approximately the same as thin vortex filaments in a classical perfect fluids. A special place is occupied by the vortex structures in the Bose\,-- Einstein condensate obtained for ultracold atomic gases \cite{fett2009}. This article will be concerned with a mathematical model of the dynamics of two vortex filaments in a Bose\,-- Einstein condensate enclosed in a harmonic trap  \cite{kevrekPhysLett2011}. This model leads to a completely Liouville integrable Hamiltonian system with two degrees of freedom, and for this reason, topological methods used in such systems can be applied. Topological methods were successfully used for investigation of the stability problem of absolute and relative choreographies \cite{borkil2000}, \cite{bormamMatMet}, \cite{bormamkil2004}, \cite{kilinbormam2013}, \cite{BorSokRyab2016}. These motions in integrable models, as a rule, correspond to the values of the constant first integrals, for which the integrals, considered as functions of phase variables, turn out to be dependent in the sense of the linear dependence of the differentials. The main role in the study of such dependence is played by the bifurcation diagram of the momentum map.

This publication is devoted to the integrable perturbation of the model considered in \cite{SokRyabRCD2017}, \cite{sokryab2018}. In this paper, the bifurcation diagram is explicitly determined and bifurcations of Liouville tori are investigated. In comparison with the results of \cite{SokRyabRCD2017}, \cite{sokryab2018}, \cite{RyabDan2019}, \cite{RyabSocND2019} we obtained expressions for the case of arbitrary signs of the vortex intensities.

\section{Model and Definitions}
Here we following the original works \cite{kevrekPhysLett2011}, \cite{navarro},
\cite{koukoul} in the description of the model.
Let us consider $N$ interacting vortices in a Bose-Einstein condensate enclosed in a harmonic trap and let $(x_k,y_k)$ is the position of the $k$-th vortex, $r_k=\sqrt{x_k^2+y_k^2}$.  A single vortex $(x_k,y_k)$ in a harmonic trap is well known to precess around the center of the trap with  the frequency $\omega_{\rm{pr}}$ which can be approximated by
$\omega_{\textrm{pr}}=\omega_{\textrm{pr}}^{0}/(1-r_k^2/R_{\textrm TF}^2)$, where the frequency at the
trap center is $\omega_{\textrm{pr}}^{0}=\ln\bigl(A\frac{\mu}{\Omega}\bigr)/R_{\textrm{TF}}^2$, $\mu$ is the chemical potential, $R_{\textrm{TF}}=\sqrt{2\mu}/\Omega$ is the so-called Thomas-Fermi~(TF) radius, $A=2\sqrt{2}\pi$ is a numerical constant, and $\Omega=\omega_r/\omega_z$, here $\omega_r$ and $\omega_z$ are the confining radial and axial frequencies of the harmonic trap, respectively. On the other hand, in the absence of a harmonic trap, two interacting vortices will rotate around each other with a frequency of $\omega_{\textrm{vort}} = B/r_{kj}^2$, where $r_{kj}=\sqrt{(x_k-x_j)^2+(y_k-y_j)^2}$ is the distance between the vortices and $B$ is a constant factor. If $(x_k,y_k)$ is the position of the $k$-th vortex, the corresponding the mathematical model of the dynamics of $N$ interacting vortices in a Bose-Einstein condensate (BEC) enclosed in a harmonic trap is described by following the system of differential equations \cite{kevrekPhysLett2011}, \cite{navarro},
\cite{koukoul}:
\begin{equation}
\label{d0}
\begin{array}{l}
\displaystyle{\dot x_k=-\Gamma_k\omega_{\textrm{pr}}y_k-\frac{B}{2}\sum\limits_{j\neq k}^{N}\,\Gamma_j\frac{y_k-y_j}{r_{kj}^2},}\\[3mm]
\displaystyle{\dot y_k=\Gamma_k\omega_{\textrm{pr}}x_k+\frac{B}{2}\sum\limits_{j\neq k}^{N}\,\Gamma_j\frac{x_k-x_j}{r_{kj}^2},}
\end{array}
\end{equation}
where $\Gamma_k$ is the charge of the $k$-th vortex, ($k=1,\ldots,N$) and $N$ is the total
number of interacting vortices.

For convenience, following to the paper \cite{koukoul} we can further rescale time to the period of the single vortex
precessing near the center of the trap ($\tau=t\omega_{\textrm pr}^0$) and pass to dimensionless variables
using the relations
\begin{equation}
\label{d1}
x_k=\tilde{x}_kR_{\textrm TF},\quad y_k=\tilde{y}_kR_{\textrm TF}.
\end{equation}

After rescaling \eqref{d1}, the equations of motion \eqref{d0} are written as
\begin{equation}
\label{d2}
\begin{array}{l}
\displaystyle{x_k^\prime=-\Gamma_k\frac{y_k}{1-r_k^2}-c\sum\limits_{j\neq k}^{N}\,\Gamma_j\frac{y_k-y_j}{r_{kj}^2},}\\[3mm]
\displaystyle{y_k^\prime=\Gamma_k\frac{x_k}{1-r_k^2}+c\sum\limits_{j\neq k}^{N}\,\Gamma_j\frac{x_k-x_j}{r_{kj}^2},}
\end{array}
\end{equation}
where the dimensionless parameter $c$ is defined by the formula
\begin{equation*}
c=\frac{B}{2\ln\bigl(A\frac{\mu}{\Omega}\bigr)}
\end{equation*}
and the prime $()^\prime$ in \eqref{d2} stands for $\tfrac{d}{d\tau}$.

The equations of motion \eqref{d2} can be represented in Hamiltonian form
\begin{equation}
\label{x1}
\Gamma_k x_k^\prime=\frac{\partial H}{\partial y_k}, \quad \Gamma_k y_k^\prime=-\frac{\partial H}{\partial x_k},\quad k=1,\ldots,N.
\end{equation}
with Hamiltonian
\begin{equation}
\label{x0}
\displaystyle{H=\frac{1}{2}\sum\limits_{k=1}^N\,\Gamma_k^2\ln(1-r_k^2)-\frac{c}{2}\sum\limits_{k=1}^N\sum\limits_{j<k}^N\,
\Gamma_k\Gamma_j\ln(r_{kj}^2).}
\end{equation}

In this work, however, we will consider a generalized mathematical model that describes not only the dynamics of two ($N =2 $) vortices in a Bose-Einstein condensate enclosed in a harmonic trap, but also the dynamics of two vortex filaments in an ideal fluid, li\-mi\-ted cy\-lin\-dri\-cal re\-gion. This model also leads to a completely Liouville integrable Ha\-mil\-to\-nian system, which is described by the following system of differential equations.


\begin{equation}
\label{eq1_1}
\displaystyle{\Gamma_k\dot x_k=\frac{\partial H}{\partial y_k} (z_1,z_2);\quad \Gamma_k\dot y_k=-\frac{\partial H}{\partial x_k} (z_1,z_2),\quad k=1,2,}
\end{equation}
where the Hamiltonian $H$ has the form
\begin{equation}
\label{eq1_2}
\begin{array}{l}
\displaystyle{H=\frac{1}{2}\Bigl[\Gamma_1^2\ln(1-|z_1|^2)+\Gamma_2^2\ln(1-|z_2|^2)+\Gamma_1\Gamma_2\ln\left(\frac{[|z_1-z_2|^2+(1-|z_1|^2)(1-|z_2|^2)]^\varepsilon}{|z_1-z_2|^{2(c+\varepsilon)}}\right)\Bigr].}
\end{array}
\end{equation}
Here, the Cartesian coordinates of $k$-th vortex ($k=1,2$) with intensities $\Gamma_k$ are denoted by\linebreak
$z_k=x_k+{\rm i}y_k$. Physical parameter ``$c$'' expresses the extent of the vortices' interaction,
$\varepsilon$ is a parameter of deformation. These parameters determine two limiting cases, namely, the model of two point vortices enclosed in a harmonic trap in a Bose-Einstein condensate ($\varepsilon=0$) \cite{kevrekPhysLett2011}, \cite{navarro}, \cite{koukoul} and the model of two point vortices bounded by a circular region in an ideal fluid ($c=0$, $\varepsilon=1$)  \cite{greenh}, \cite{bormamMatMet}.
The phase space $\cal P$ is defined as a direct product of two open disks of radius $1$ with the exception of vortices' collision points
\begin{equation*}
{\cal P}=\{(z_1,z_2)\,:\, |z_1|<1,\,|z_2|<1,z_1\ne z_2\}.
\end{equation*}
 The Poisson structure on the phase space $\cal P$ is given in the standard form
\begin{equation}
\label{eq1_3}
\{z_k,\bar{z}_j\}=-\frac{2\rm i}{\Gamma_k}\delta_{kj},
\end{equation}
 where $\delta_{kj}$ is the Kronecker delta.

The system $\eqref{eq1_1}$ admits an additional first integral of motion, \textit{the angular momentum of vorticity},
\begin{equation}
\label{eq1_4}
F=\Gamma_1|z_1|^2+\Gamma_2|z_2|^2.
\end{equation}

The function $F$ together with the Hamiltonian $H$ forms on $\cal P$ a complete involutive set of integrals of system $\eqref{eq1_1}$. According to the Liouville-Arnold theorem, a regular level surface of the first integrals is a nonconnected union of two-dimensional tori filled with conditionally periodic trajectories. The \textit{momentum map} ${\cal F}\,:\, {\cal P}\to {\mathbb R}^2$ is defined by setting  ${\cal F}(\boldsymbol x)=(F(\boldsymbol x), H(\boldsymbol x))$. Let $\cal C$ denote the set of all critical points of the momentum map, i.e., points at which $\mathop{\rm rank}\nolimits d{\cal F}(\boldsymbol x) < 2$. The set of critical values $\Sigma = {\cal F}({\cal C}\cap{\cal P})$ is called the \textit{bifurcation diagram}.

In works \cite{SokRyabRCD2017} and \cite{sokryab2018} the bifurcation diagram was analytically investigated at $c=1$ and $\varepsilon=0$. In \cite{RyabDan2019} and \cite{RyabSocND2019} a reduction to a system with one degree of freedom was performed and a bifurcation of three tori into one was found at $c>3$ and $\varepsilon=0$.
This bifurcation was observed earlier by Kharlamov \cite{Kharlamov1988} while studying a phase topology of the Goryachev-Chaplygin-Sretensky integrable case in rigid body dynamics. In Fomenko, Bolsinov, and Matveev's work \cite{bolsmatvfom1990} it was found as a singularity in a 2-atom form of a Liouville foliation's singular layer. In Oshemkov and Tuzhilin's work \cite{oshtuzh2018}, devoted to the splitting of saddle singularities, such a bifurcation was found to be unstable and its perturbed foliations were presented. In the situation where the physical parameter of vortices' intensity ratio is experiencing integrable perturbation, said bifurcation comes down to the bifurcation of two tori into one and vice versa \cite{RyabDan2019}. In another limiting case ($c=0, \varepsilon=1$), the bifurcation analysis of dynamics of two point vortices bounded by a circular domain in an ideal fluid is performed \cite{bormamMatMet}. In these limiting cases completely different bifurcation diagrams were obtained. In the case of a positive vortex pair a new bifurcation diagram is obtained for which the bifurcation of four tori into one is observed \cite{RyabovDanPhys2019}. The presence of three-into-one and four-into-one tori bifurcations in the integrable model of vortex dynamics with positive intensities indicates a complex transition and connection between two bifurcation diagrams in both limiting cases.
In this paper we analytically derive the equations that define a parametric family of bifurcation diagrams of the generalized model \eqref{eq1_1} containing bifurcation diagrams of the specified limiting cases. In the general case reduction to a system with one degree of freedom allows us to apply level curves of corresponding Hamiltonian in order to observe different kinds of Liouville tori bifurcations.

\section{Bifurcation diagram}
We define the polynomial expressions $F_1$ and $F_2$ from phase variables:
\begin{eqnarray}
&\label{eq2_1}
F_1=x_1y_2-y_1x_2,\\[3mm]
&
\begin{array}{l}
\label{eq2_2}
F_2= cx_2(\Gamma_1x_1+\Gamma_2x_2)(x_2^2+y_2^2-1)[x_2(x_1x_2-1)+x_1y_2^2][(x_1^2-1)x_2^2+x_1^2y_2^2]+\\[3mm]
+\Gamma_2[(x_1^2-1)x_2^2+x_1^2y_2^2]\Bigl\{\varepsilon x_2^3(x_2^2+y_2^2-1)^2+x_1(x_1-x_2)(x_2^2+y_2^2)[x_2(x_1x_2-1)+x_1y_2^2]\Bigr\}\\[3mm]
+\Gamma_1x_1(x_2^2+y_2^2-1)\Bigl\{x_2^2(x_2-x_1)(x_2^2+y_2^2)[x_2(x_1x_2-1)+x_1y_2^2]+\varepsilon[(x_1^2-1)x_2^2+x_1^2y_2^2]^2\Bigr\},
\end{array}
\end{eqnarray}
and denote by ${\cal N}$ the closure of system's set of solutions:
\begin{equation}
\label{eq2_3}
F_1 = 0,\quad F_2 = 0.
\end{equation}

Then the theorem below is true.
\begin{theorem}
The set of critical points $\cal C$ of the momentum map $\cal F$ coincides with the set of solutions for the system \eqref{eq2_3}. The set ${\cal N}$ is a two-dimensional invariant submanifold of the system \eqref{eq1_1} with the Hamiltonian \eqref{eq1_2}.
\end{theorem}

\proof
To prove the first statement of the theorem it is necessary to find the
phase space points where the rank of the momentum map is not maximal. With the help of direct computations one can verify that the Jacobi matrix of the momentum map has
zero minors of the second order at the points ${\boldsymbol z}\in{\cal P}$, the coordinates of which satisfy the equations of system \eqref{eq2_3}.
Therefore ${\cal C} = {\cal N}$. The fact that the relations \eqref{eq2_3} are invariate might be prooved by the following chain of correct equalities:
\begin{equation*}
\dot F_1 = \{F_1,H\}_{F_1=0} = \sigma_1F_2,\quad \dot F_2 = \{F_2,H\}_{F_1=0} = -\frac{x_2y_2}{x_2^2+y_2^2}\sigma_1\sigma_2F_2,
\end{equation*}
where polynomial functions $\sigma_k$ from phase variables have the following explicit form:
\begin{equation*}
\begin{array}{l}
\displaystyle{\sigma_1=\frac{1}{(x_1 - x_2)x_2(x_2^2 + y_2^2-1)[x_2(x_1x_2-1) + x_1y_2^2][(x_1^2-1)x_2^2 + x_1^2y_2^2]},}\\[3mm]
\end{array}
\end{equation*}
\begin{equation*}
\begin{array}{l}
\sigma_2=c(x_2^2+y_2^2-1)\Bigl\{\Gamma_1\Bigl[x_2^3\Bigl(4+x_1\bigl((x_1^2-3)x_2-2x_1\bigr)\Bigr)+x_1x_2\bigl((2x_1^2-3)x_2-2x_1\bigr)y_2^2+x_1^3y_2^4\Bigr]-\\[3mm]
-\Gamma_2x_2(x_2^2+y_2^2)(-2x_1x_2-x_2^2+3x_1^2(x_2^2+y_2^2))\Bigr\}-\Gamma_2(x_2^2+y_2^2)\Bigl\{4x_2^3+x_1^3(x_2^2+y_2^2)(1+x_2^2+y_2^2)+\\[3mm]
+x_1x_2^2(2\varepsilon(x_2^2+y_2^2-1)^2-3(1+x_2^2+y_2^2))\Bigr\}-\Gamma_1x_2(x_2^2+y_2^2-1)\Bigl\{4\varepsilon[(x_1^2-1)x_2^2+x_1^2y_2^2]+\\[3mm]
+(x_2^2+y_2^2)[2x_1x_2(1+x_2^2+y_2^2)-x_2^2-3x_1^2(x_2^2+y_2^2)]\Bigr\}.
\end{array}
\end{equation*}
\qed

To determine the bifurcation diagram $\Sigma $ as the image of the set $\cal C $ of critical points
of the momentum map $\cal F $, it is convenient to change to polar coordinates
\begin{equation}
\label{eq3_1}
x_1 = r_1\cos\theta_1,\quad y_1 = r_1\sin\theta_1,\quad
x_2 = r_2\cos\theta_2,\quad y_2 = r_2\sin\theta_2.
\end{equation}
Substitution of \eqref{eq3_1} into the first equation of the system \eqref{eq2_3} results in an equation $\sin(\theta_1-\theta_2)=0$, i.e. $\theta_1-\theta_2=0$ or $\theta_1-\theta_2=\pi$.

In case of $\theta_1=\theta_2+\pi$, the second equation of the system \eqref{eq2_3} is reduced to

\begin{equation}
\label{eq3_2}
W_1(r_1,r_2)=0,
\end{equation}
where
\begin{equation*}
\begin{array}{l}
W_1(r_1,r_2)=(1-r_1^2)(1-r_2^2)\Bigl\{[c(1+r_1r_2)+\varepsilon](\Gamma_1r_1-\Gamma_2r_2)-\varepsilon(\Gamma_1r_1^3-\Gamma_2r_2^3)\Bigr\}-\\[3mm]
-r_1r_2(r_1+r_2)(1+r_1r_2)[\Gamma_1(1-r_2^2)-\Gamma_2(1-r_1^2)].
\end{array}
\end{equation*}

In case of $\theta_1=\theta_2$, the second equation of the system \eqref{eq2_3} is reduced to
\begin{equation}
\label{eq3_21}
W_2(r_1,r_2)=0,
\end{equation}
where
\begin{equation*}
\begin{array}{l}
W_2(r_1,r_2)=W_1(r_1,-r_2).
\end{array}
\end{equation*}

Substituting \eqref{eq3_1} into the Hamiltonian \eqref{eq1_2} and the vorticity momentum \eqref{eq1_4} in the case where $\theta_1=\theta_2+\pi$, leads to the following values of the first integrals:
\begin{equation}
\label{eq3_3}
\gamma_1:\left\{
\begin{array}{l}
\displaystyle{h_1(r_1,r_2)=\frac{1}{2}\left\{\Gamma_1^2\ln(1-r_1^2)+\Gamma_2^2\ln(1-r_2^2)+\Gamma_1\Gamma_2\ln\Bigl[\frac{(1+r_1r_2)^{2\varepsilon}}{(r_1+r_2)^{2(c+\varepsilon)}}\Bigr]\right\},}\\[3mm]
f=\Gamma_1r_1^2+\Gamma_2r_2^2.
\end{array}\right.
\end{equation}
In case of $\theta_1=\theta_2$,
\begin{equation}
\label{eq3_31}
\gamma_2:\left\{
\begin{array}{l}
\displaystyle{h_2(r_1,r_2)=h_1(r_1,-r_2),}\\[3mm]
f=\Gamma_1r_1^2+\Gamma_2r_2^2.
\end{array}\right.
\end{equation}

These systems \eqref{eq3_3}, \eqref{eq3_31} together with the equations \eqref{eq3_2}, \eqref{eq3_21} defines an implicit bifurcation diagram on the plane ${\mathbb R}^2(f,h)$.

The bifurcation diagram for a system of two vortices of opposite signs, corresponding to the following parameter values ($N = 2 $, $ \Gamma_1 = 1 $, $ \Gamma_2 = -0.45$, $c = 1.5$, $\varepsilon = 3 $) is shown in Fig.~\ref{fig1}. The following values of the constants of the first integrals are given here $h_0=-1.75$, $h_1=-1.3$, $h_d=-1.9791878639$, $f_a=0.05$, $f_b=0.0847786118$ (point $b\in\gamma_2$), $f_c=0.09$, $f_e=0.198$, $f_d=0.1537691986$ (point $d\in\gamma_2\cap\gamma_3$),  $f_1=-0.15$, $f_2=-0.07707$, $f_3=-0.075$. This is a diagram similar to the one constructed in \cite{SokRyabRCD2017}.

\begin{figure}[!ht]
  \centering
  \includegraphics[width=0.8\textwidth]{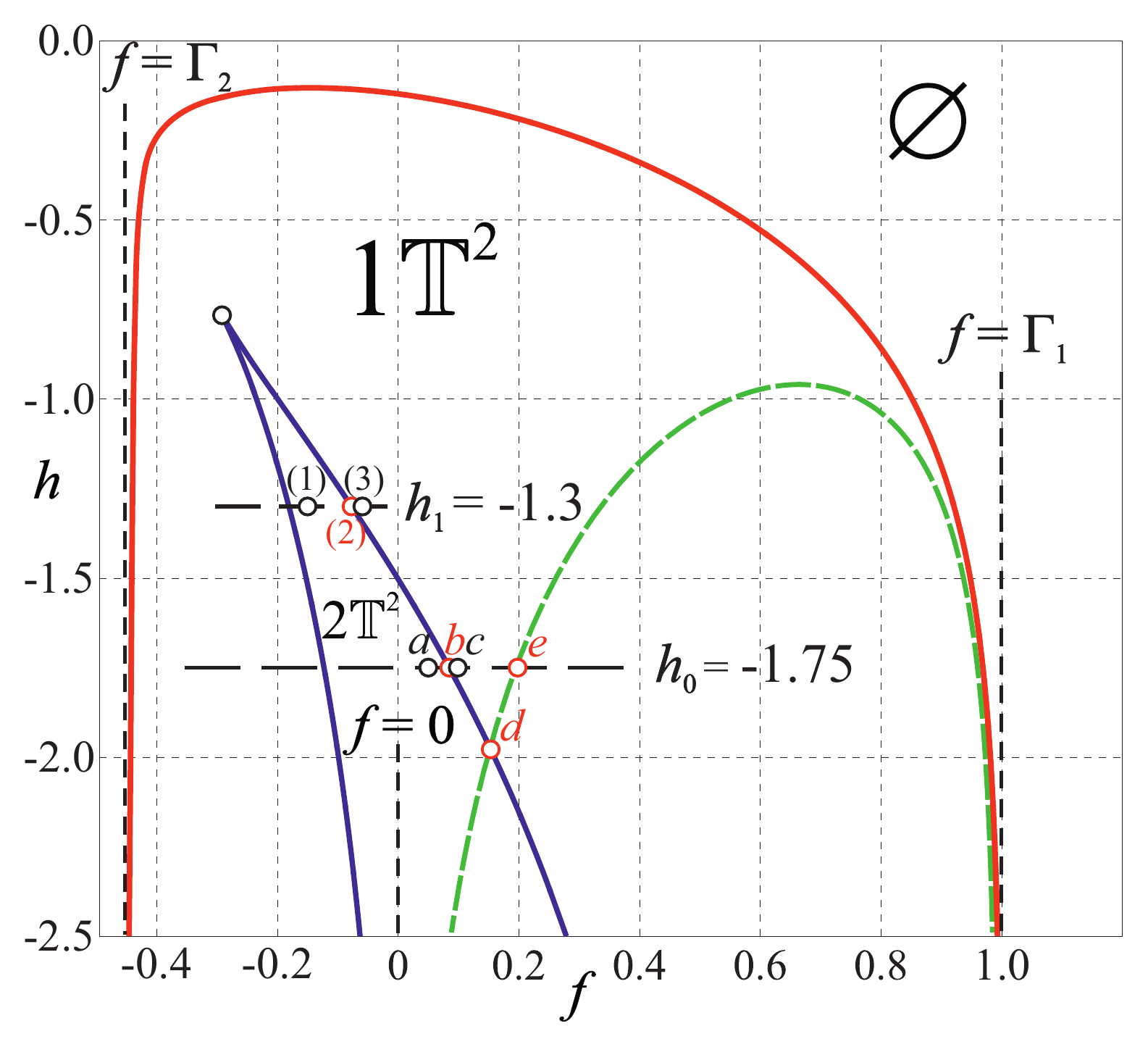}
  \caption{Bifurcation diagram for the parameters $\Gamma_1=1$, $\Gamma_2=-0.45$, $c=1.5$, $\varepsilon=3$.}
\label{fig1}
\end{figure}

In this paper, we define a bifurcation diagram as a two-parameter family. Figure~\ref{fig2} shows the new type of the bifurcation diagram from this family, which was not observed in any of the works \cite{SokRyabRCD2017}, \cite{RyabShad}, \cite{RyabSocND2019}. Such a diagram contains two ``loops'' with the cusped points, and one of the loops has only one vertical asymptote. The figure also shows a separating curve $\gamma_3$ that is responsible for changing the projection of the torus, without changing their number. The equation of this curve is given explicitly in the following form:
\begin{equation*}
\gamma_3:\, h=\frac{1}{2}\left\{\Gamma_1^2 \ln\left(1-\frac{f}{\left|\Gamma_1\right|}\right)-\Gamma_1\Gamma_2(c+\varepsilon) \ln\left(\frac{f}{\left| \Gamma_1\right| }\right)\right\}\qquad 0<f<\left|\Gamma_1\right|.
\end{equation*}

\begin{figure}[!ht]
  \centering
  \includegraphics[width=1\textwidth]{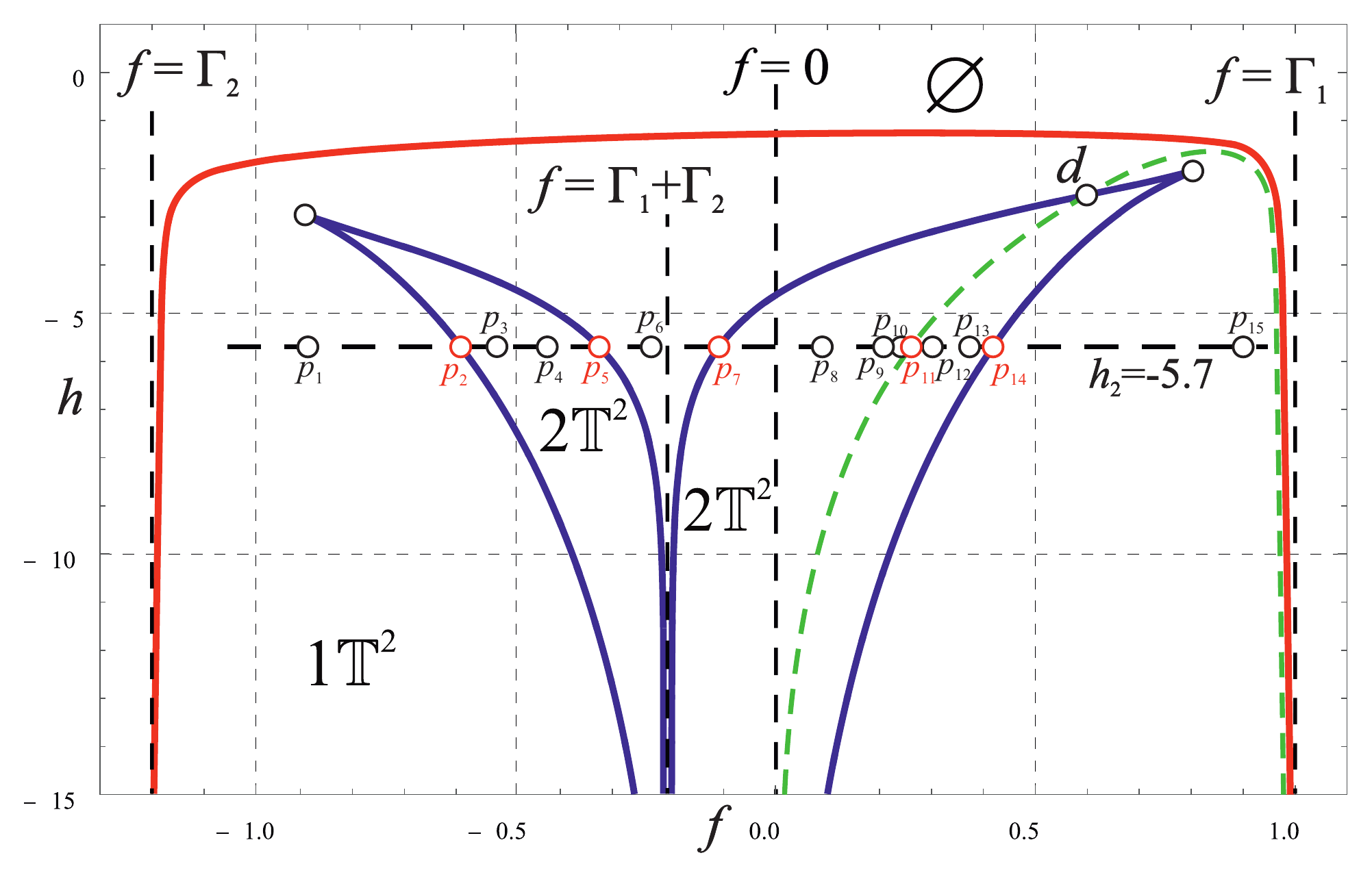}
  \caption{New bifurcation diagram for the parameters $\Gamma_1=1$, $\Gamma_2=-1.2$, $c=0.09$, $\varepsilon=7$.}
\label{fig2}
\end{figure}

The values of the second integral $f=p_k, k=1,\ldots,15$ along the line $h = h_2=-5.7$ are presented in the Table~\ref{table1}, where
the points $p_2,p_5,p_7, p_{14}\in\gamma_2$, point $d\in\gamma_2\cap\gamma_3$, point $p_{11}\in{(h=h_2)\cap\gamma_3}$.

\begin{table}
\caption{The values of the constant of the second integral  $f=p_k$.}
\begin{tabular}{|c|c|c|c|c|c|c|c|c|c|}
\hline
$p_1$  &$p_2$        &  $p_3$  & $p_4$   & $p_5$      & $p_6$   &$p_7$       &$p_8$  &$p_9$    &$p_{10}$\\ \hline
$-0.9$ & $-0.591655$ & $-0.59$ &$-0.425$ &$-0.324902$ &$-0.225$ &$-0.108766$ &$0.09$ &$0.2685$ &$0.25$ \\ \hline
\end{tabular}
\\[3mm]
\begin{tabular}{|c|c|c|c|c|}
\hline
$p_{11}$    &$p_{12}$&$p_{13}$ &$p_{14}$ &$p_{15}$\\ \hline
$0.271816$ &$0.35$  &$0.415$  &$0.418417$ & $0.9$\\ \hline
\end{tabular}
\label{table1}
\end{table}

\section{Reduction to a system with one degree of freedom}
In the works \cite{RyabSocND2019}, \cite{RyabShad} bifurcation diagrams were obtained in the case of two identical vortices and vortices with intensities of the same signs, respectively. In the previous sections of this work, the presentation did not imply any restrictions on the signs of the vortex intensities. Throughout what follows, we assume that vortices have intensities of $\Gamma_1$ and $\Gamma_2$ of \textit{any} signs. Let's perform an explicit reduction to a system with one degree of freedom. In order to perform this for the system \eqref{eq1_1} with Hamiltonian \eqref{eq1_2}, one should substitute phase variables $(x_k, y_k)$ to new variables $(u,v,\alpha)$ using the formulas below

\begin{equation*}
\begin{array}{l}
\displaystyle{x_{1} =\frac{\sgn(\Gamma_{1}\Gamma_{2})}{\sqrt{|\Gamma _{1} |}}[ u\cos( \alpha ) -v\sin( \alpha )],}\\[5mm]
\displaystyle{y_{1} =\frac{\sgn(\Gamma_{1}\Gamma_{2})}{\sqrt{|\Gamma _{1} |}}[ u\sin( \alpha ) +v\cos( \alpha )],}\\[5mm]
\displaystyle{x_{2} =\frac{\cos( \alpha )}{\sqrt{|\Gamma_{2} |}}\sqrt{\sgn(\Gamma_{1}\Gamma_{2})\left[ f\,\sgn \Gamma_{1}-\left( u^{2} +v^{2}\right)\right]},}\\[5mm]
\displaystyle{y_{2} =\frac{\sin( \alpha )}{\sqrt{|\Gamma_{2} |}}\sqrt{\sgn(\Gamma_{1}\Gamma_{2})\left[ f\,\sgn \Gamma_{1}-\left( u^{2} +v^{2}\right)\right]}.}
\end{array}
\end{equation*}

The physical variables $(u,v)$ are Cartesian coordinates of one of the vortices in a coordinate system that is associated with another vortex rotating around the center of vorticity. The choice of such variables is suggested by the presence of the integral of the angular momentum of vorticity \eqref{eq1_4}, which is invariant under the rotation group $SO(2)$. The existence of a one-parameter symmetry group allows to perform a reduction to a system with one degree of freedom in a similar fashion as in mechanical systems with symmetry \cite{Kharlamov1988}. Backward substitution

\begin{equation*}
\displaystyle{U=\frac{\sqrt{|\Gamma _{1} |}( x_{1} x_{2} +y_{1} y_{2})}{\sgn(\Gamma_{1}\Gamma_{2})\sqrt{x^{2}_{2} +y^{2}_{2}}} ,\quad V=\frac{\sqrt{|\Gamma _{1} |}( x_{2} y_{1} -x_{1} y_{2})}{\sgn(\Gamma_{1}\Gamma_{2})\sqrt{x^{2}_{2} +y^{2}_{2}}}}
\end{equation*}
leads to canonical variables with respect to the bracket \eqref{eq1_3}:
\begin{equation*}
\{U,V\}=-\{V,U\}=\sgn\Gamma_1,\quad \{U,U\}=\{V,V\}=0.
\end{equation*}
The system with respect to the variables $(u,v)$ is Hamiltonian:
\begin{equation}\label{z2}
\dot u=\frac{\partial H_1}{\partial v}\sgn\Gamma_1,\quad
\dot v=-\frac{\partial H_1}{\partial u}\sgn\Gamma_1,
\end{equation}
with Hamiltonian
\begin{equation}\label{z3}
 \begin{array}{l}
\displaystyle{H_1=\frac{1}{2}\left\{\Gamma^{2}_{1}\ln\left(1-\frac{u^{2} +v^{2}}{|\Gamma_{1} |}\right) +
\Gamma^{2}_{2}\ln\left( 1-\frac{\sgn(\Gamma_{1}\Gamma_{2})\left[f\sgn\Gamma_{1}-(u^2+v^2)\right]}{|\Gamma_{2} |}\right)-\right.}
\\[3mm]
\displaystyle{-\Gamma_{1}\Gamma_{2}( c+\varepsilon )\ln\left[\left(\frac{\sgn\Gamma_1 u}{\sqrt{|\Gamma_{1}|}}-\frac{\sgn\Gamma_2\sqrt{\sgn(\Gamma_{1}\Gamma_{2})\left[f\sgn \Gamma_{1}-\left(u^{2}+v^{2}\right)\right]}}{\sqrt{|\Gamma_{2}|}}\right)^2+\frac{v^{2}}{|\Gamma_{1}|}\right]+}\\[3mm]
\displaystyle{+\varepsilon\Gamma_{1}\Gamma_{2}\ln\left[
\left(1-\frac{\sgn(\Gamma_{1}\Gamma_{2})u\sqrt{\sgn(\Gamma_1\Gamma_{2})\left[f\sgn\Gamma_{1}-\left(u^{2}+v^{2}\right)\right]}}{\sqrt{|\Gamma _{1}||\Gamma_{2}|}}\right)^2+\right.}\\[3mm]
\displaystyle{\left.\left.+\frac{\sgn(\Gamma_1\Gamma_{2}) v^{2}\left[f\sgn\Gamma_{1}-\left(u^{2}+v^{2}\right)\right]}{|\Gamma _{1} ||\Gamma _{2} |}\right]\right\}.}
\end{array}
\end{equation}

The rotation angle $\alpha(t)$ of the rotating coordinate system satisfies the differential equation
\begin{equation*}
\begin{array}{l}
\displaystyle{\dot{\alpha} =\frac{\Gamma_2^2}{\Gamma_2-f+\sgn\Gamma_1 \left(u^2+v^2\right)}+
c\frac{\Gamma_1 |\Gamma_2|\sqrt{|\Gamma_1|}}{\sgn(\Gamma_{1}\Gamma_{2})}\frac{R_1(u,v)}{Q_1(u,v)}+}\\[5mm]
\displaystyle{+\varepsilon\Gamma_1|\Gamma_2|\sqrt{|\Gamma_2|}\frac{|\Gamma_1|-u^2-v^2}{\sqrt{\sgn(\Gamma_1\Gamma_2)
\left[f\sgn\Gamma_{1}-\left(u^{2}+v^{2}\right)\right]}}\frac{R_2(u,v)}{Q_2(u,v)}},
\end{array}
\end{equation*}
where
\begin{equation*}
\begin{array}{l}
\displaystyle{R_{1}(u,v) = \Gamma_{1}\left(f\sgn\Gamma_{1}-u^{2} -v^{2}\right) -\Gamma_{2}u^{2},}\\[5mm]
\displaystyle{Q_{1}(u,v) =\sqrt{|\Gamma_{2}|}u\sqrt{\sgn(\Gamma_{1}\Gamma_{2})\left(f\sgn\Gamma_{1}-u^{2}-v^{2}\right)}\times}\\[5mm]
\displaystyle{\times\left[\Gamma_{2}\left(u^{2}+v^{2}\right)-\Gamma_{1}\left(f\sgn\Gamma_{1}-u^{2}-v^{2}\right)\right]+}\\[3mm]
\displaystyle{+\sqrt{|\Gamma _{1}|}\left(f\sgn\Gamma_{1}-u^{2}-v^{2}\right)\left[\Gamma_{1}\left(f\sgn\Gamma_{1}-u^{2}-v^{2}\right)-\Gamma_{2}
\left(u^{2}-v^{2}\right)\right],}\\[5mm]
\displaystyle{R_{2}(u,v) =\sqrt{|\Gamma_{2}|}\sqrt{\sgn(\Gamma_{1}\Gamma_{2})\left(f\sgn\Gamma_{1}-u^{2}-v^{2}\right)}\left(|\Gamma_{1}|+u^{2}+v^{2}\right)-}\\[5mm]
\displaystyle{-\sgn(\Gamma_{1}\Gamma_{2})\sqrt{|\Gamma_{1}|}u\left[|\Gamma_{2}|+\sgn(\Gamma_{1}\Gamma_{2})\left(f\sgn\Gamma_{1}-u^{2}-v^{2}\right)\right],}\\[5mm]
\end{array}
\end{equation*}
\begin{equation*}
\begin{array}{l}
\displaystyle{Q_{2}(u,v) =\sgn(\Gamma_{1}\Gamma_{2})|\Gamma_{2}|\bigl[|\Gamma_{1}|\left(|\Gamma_{1}|+4u^{2}\right)+\left(u^{2}+v^{2}\right)^{2}\bigr]
\left(f\sgn\Gamma_{1}-u^{2}-v^{2}\right) +}\\[3mm]
\displaystyle{+|\Gamma_{1} |\left( u^{2} +v^{2}\right)\bigl[ \Gamma^{2}_{2} +\left( f \sgn\Gamma_{1} -u^{2}-v^{2}\right)^{2}\bigr]-}
\\[3mm]
\displaystyle{-2\,u\,\sgn(\Gamma_{1}\Gamma_{2})\sqrt{|\Gamma_{1}||\Gamma_{2}|}\sqrt{\sgn(\Gamma_{1}\Gamma_{2})
\left(f\sgn\Gamma_{1}-u^{2}-v^{2}\right)}\left(|\Gamma_{1}|+u^{2}+v^{2}\right)\times}\\[3mm]
\displaystyle{\times\left[ |\Gamma_{2} |+\sgn(\Gamma_{1}\Gamma_{2})\left( f \sgn \Gamma_{1}-u^{2}-v^{2}\right)\right]}.
\end{array}
\end{equation*}

The fixed points of the reduced system \eqref{z2} are determined by the critical points of the reduced Hamiltonian \eqref{z3} and correspond to the relative equilibria of vortices in the system \eqref{eq1_1}.
\begin{figure}[!ht]
  \centering
  \includegraphics[width=1\textwidth]{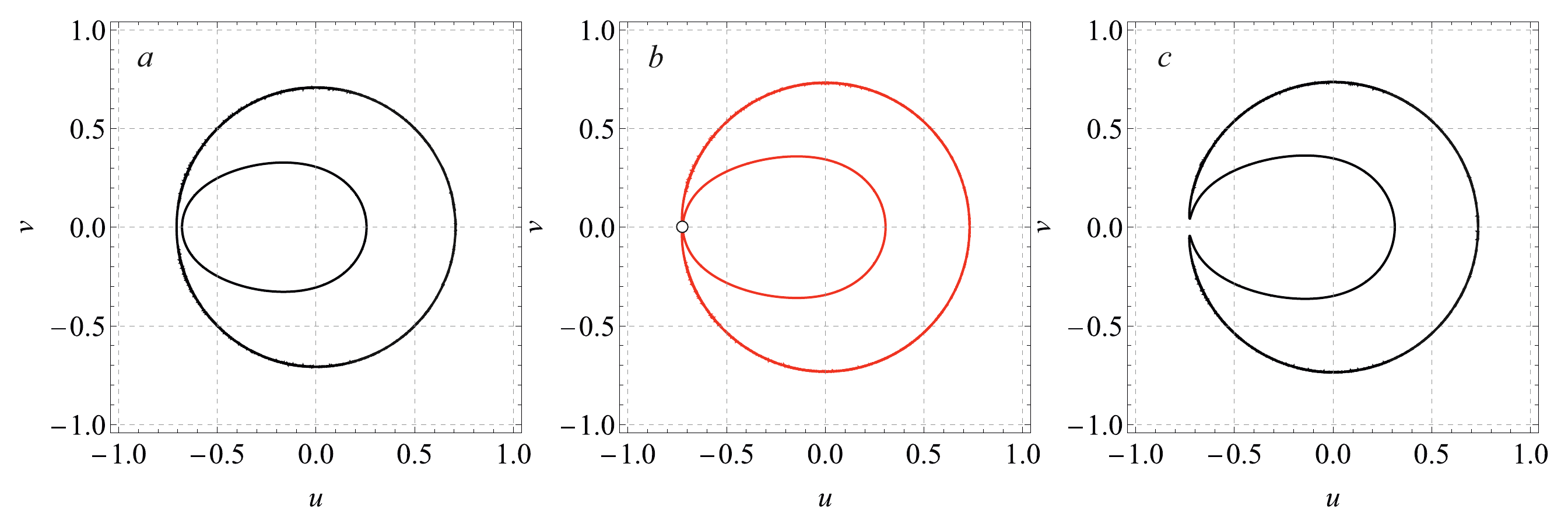}
  \caption{The level lines of the reduced Hamiltonian $H_1$ for the parameters $\Gamma_1=1$, $\Gamma_2=-0.45$, $c=1.5$, $\varepsilon=3$, corresponding to the points for $h_0=-1.75$, $f_a=0.05$, $f_b=0.0847786118$ (point $b\in\gamma_2$), $f_c=0.09$.}
\label{fig3}
\end{figure}
For a fixed value of an integral of the momentum of vorticity $f$, the regular levels of the reduced Hamiltonian are compact and motions occur along closed curves. It can be shown that the critical values of the reduced Hamiltonian determine the bifurcation diagram \eqref{eq3_2} -- \eqref{eq3_3}. Fig.~\ref{fig2} corresponds to the section of the bifurcation diagram in Fig.~\ref{fig1} along the line $h = h_0=-1.75$.
Here, for specified values of the second integral's parameter $f =f_a, f_b, f_c$, the Hamiltonian level lines of the reduced system $H_1$ is clearly presented (Fig.~\ref{fig3}). Fig.~\ref{fig4} corresponds to the section of the bifurcation diagram in Fig.~\ref{fig2} along the line $h = h_2 = -5.7$.
Fig.~\ref{fig5} shows a typical picture of the level lines of the reduced Hamiltonian $H_1$ in the case of the intersection of one of the branches of the bifurcation curve $\gamma_2$ and the separating curve  $\gamma_3$. In Fig.~\ref{fig1}, this is a point $d=(f_d = 0.1537691986, h_d =-1.9791878639)\in\gamma_2\cap\gamma_3$ for the parameters $\Gamma_1=1$, $\Gamma_2=-0.45$, $c=1.5$, $\varepsilon=3$, and in Fig.~\ref{fig2}, this is a point $d=(f_d=0.6260532,h_d=-2.4840538628)$ for the parameters $\Gamma_1=1$, $\Gamma_2=-1.2$, $c=0.09$, $\varepsilon=7$, respectively.
The Fig.~\ref{fig6} shows different types of trajectories of the vortices themselves on the plane $(x,y)$, corresponding to the points for $f_1=-0.15$;
$f_2 = -0.07707$ and $f_3 = -0.075$ of the bifurcation diagram in Fig.~\ref{fig1} along the line $h_1=-1.3$. Absolute motions of the vortices in the form of the one component torus winding  for the following values of the constants of the first integrals $h_1=-1.3$, $f_1=-0.15$, and for $h_3=-5.7$, $f=0.283$   are presented on Fig.~\ref{fig7}.

\begin{figure}[!ht]
  \centering
  \includegraphics[width=1\textwidth]{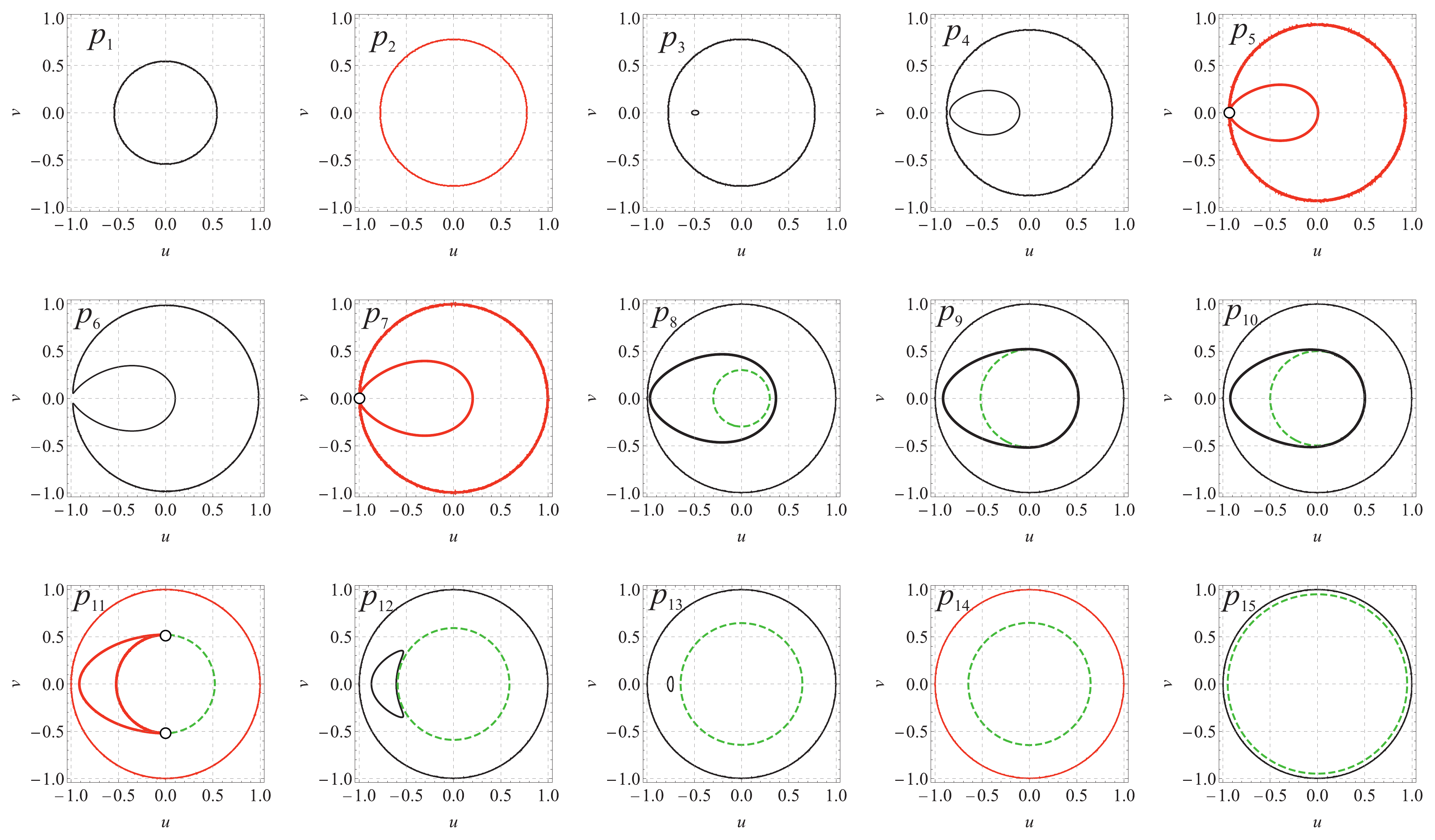}
  \caption{The level lines of the reduced  Hamiltonian $H_1$ for the parameters $\Gamma_1=1$, $\Gamma_2=-1.2$, $c=0.09$, $\varepsilon=7$, corresponding to the points for $h_2=-5.7$, $f_k=p_k, k=1,\ldots,15$.}
\label{fig4}
\end{figure}

\begin{figure}[!ht]
  \centering
  \includegraphics[width=0.32\textwidth]{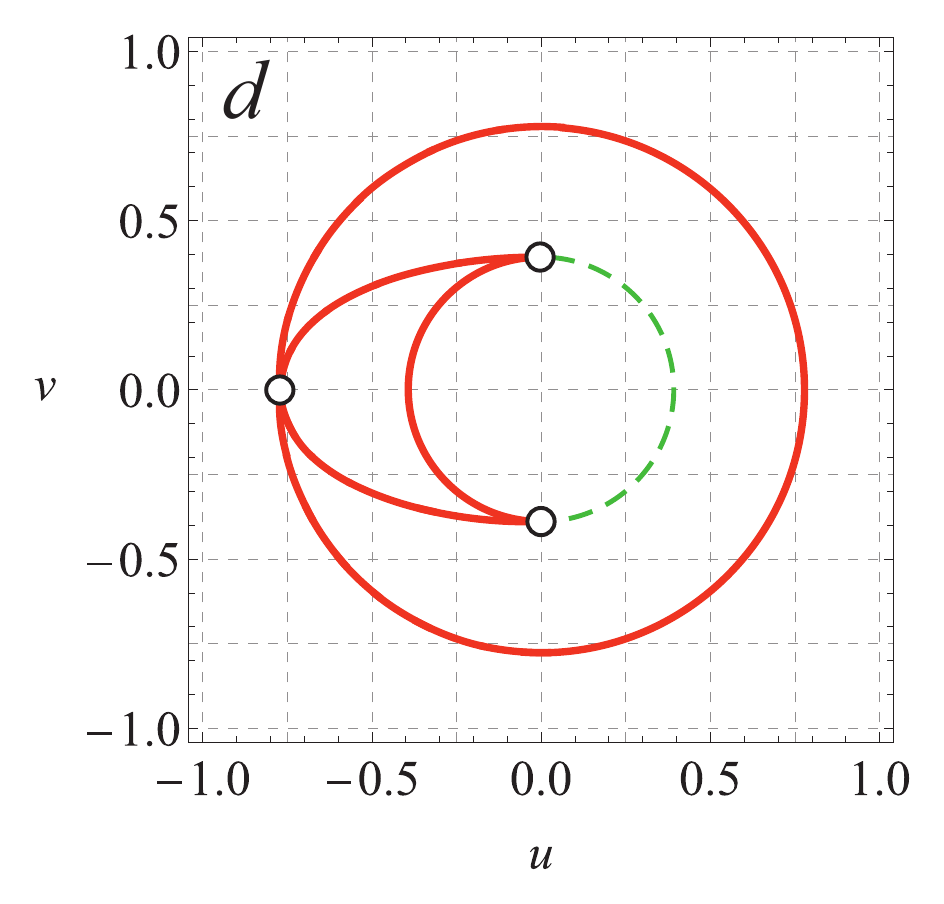}
  \includegraphics[width=0.32\textwidth]{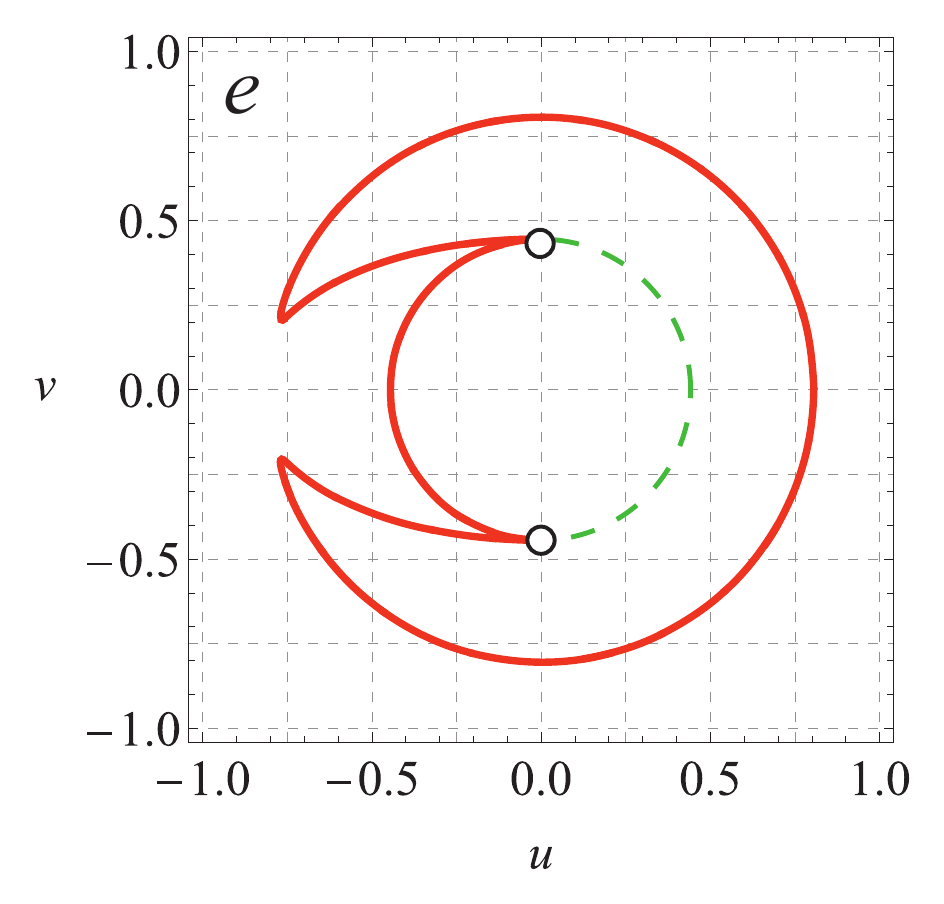}
  \includegraphics[width=0.32\textwidth]{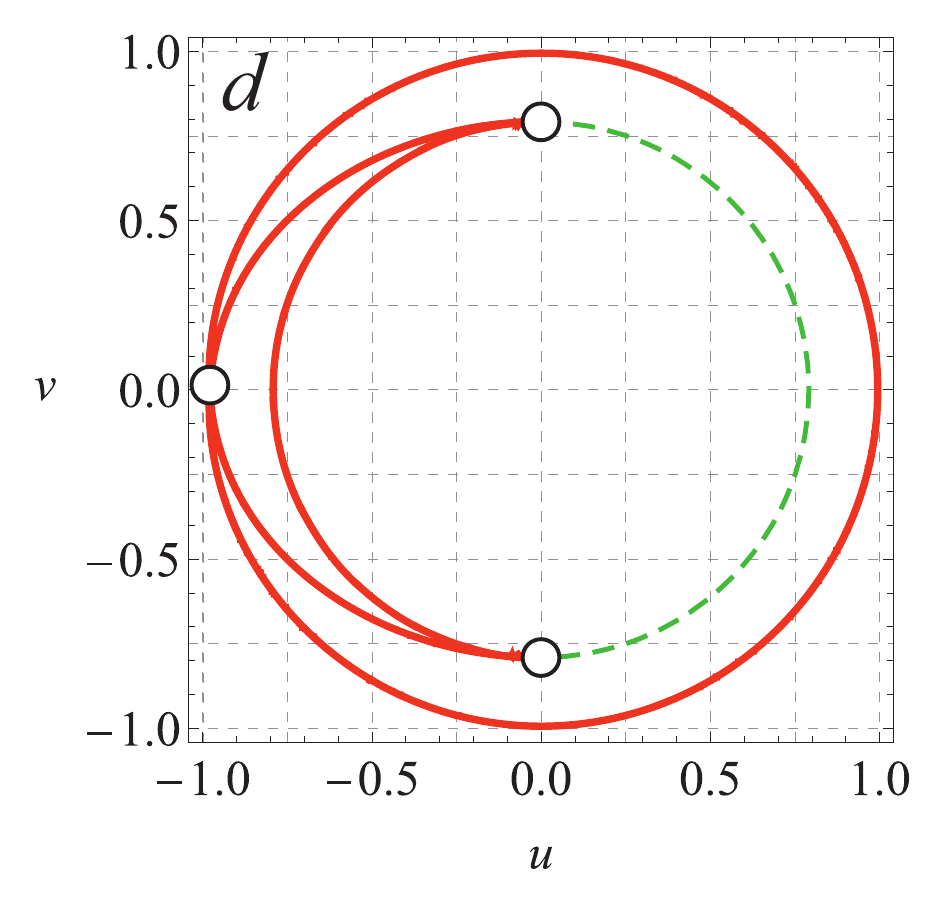}
  \caption{The level line of the reduced  Hamiltonian $H_1$ corresponding to the points  $d\in\gamma_2\cap\gamma_3$ and $e=(h=h_0)\cap\gamma_3$ in the Fig.~\ref{fig1} and Fig.~\ref{fig2}, respectively.}
\label{fig5}
\end{figure}


\begin{figure}[!ht]
  \centering
  \includegraphics[width=1\textwidth]{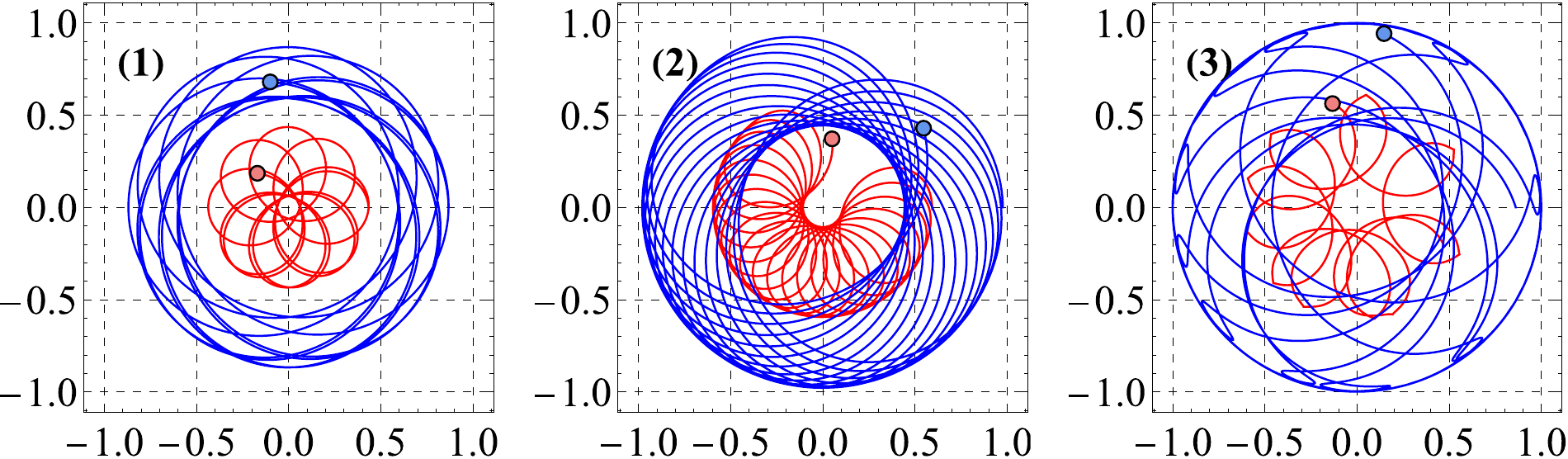}
  \caption{Vortex trajectories for the parameters $\Gamma_1=1$, $\Gamma_2=-0.45$, $c=1.5$, $\varepsilon=3$, corresponding to the points for $h_1=-1.3$, $f_1=-0.15$, $f_2=-0.07707$, $f_3=-0.075$.}
\label{fig6}
\end{figure}

\begin{figure}[!ht]
  \centering
  \includegraphics[width=0.45\textwidth]{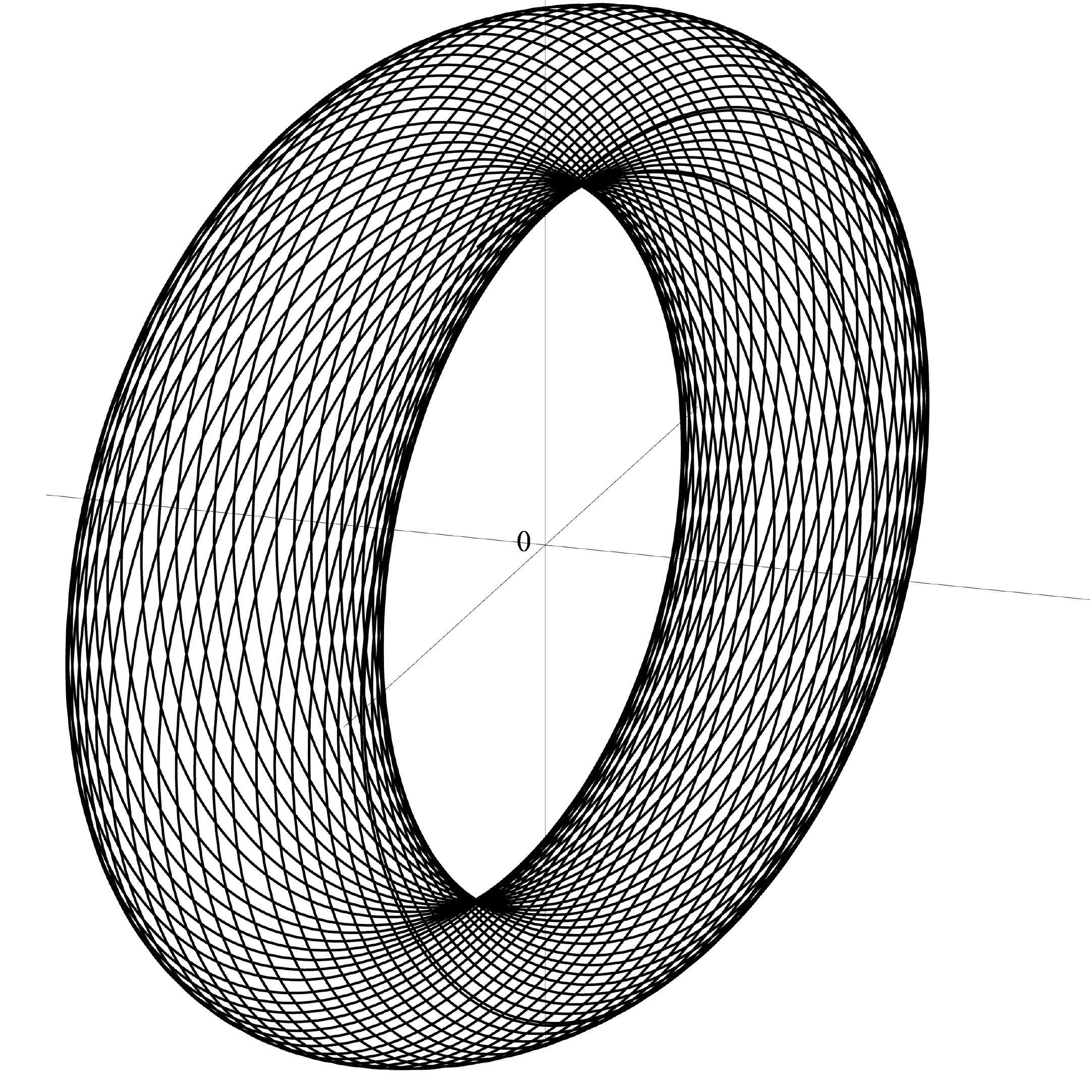}
  \includegraphics[width=0.5\textwidth]{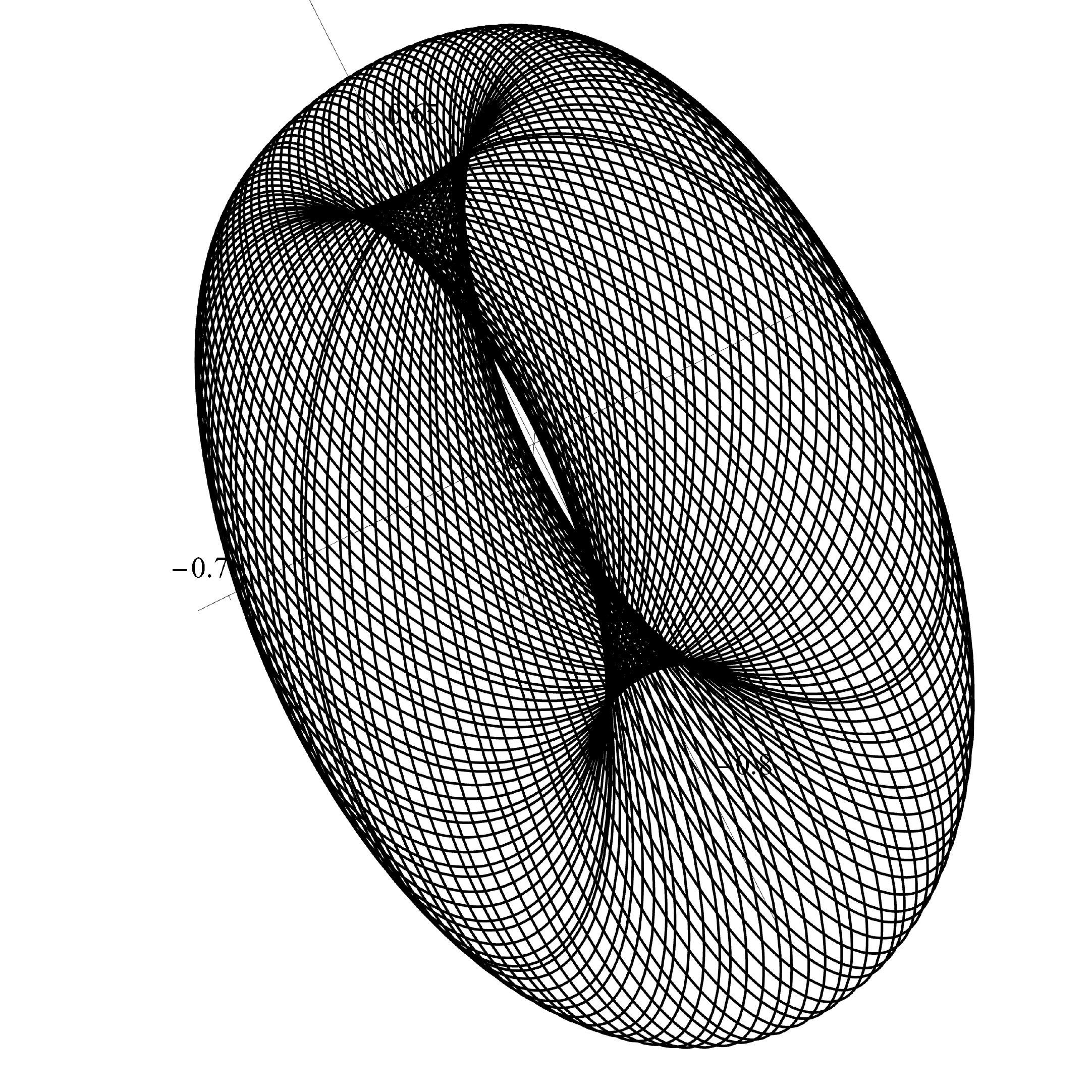}
  \caption{Absolute motions of the vortices in the form of torus winding.}
\label{fig7}
\end{figure}

\section{Conclusion}

In this paper, we derive a bifurcation diagram of the momentum map in the case of a generalized system that describes the dynamics of two vortices of opposite intensities, both in an ideal fluid placed inside a circular cylinder and in a Bose-Einstein condensate enclosed in a harmonic trap. Recall that in a case of two vortices of the same sign, considered in \cite{RyabShad}, the bifurcation diagram of vortices in a Bose-Einstein condensate differs significantly from the diagram of vortices in an ideal fluid. Here we found that, on the one hand, there are diagrams similar to those constructed in the \cite{SokRyabRCD2017}, \cite{RyabShad}, \cite{RyabSocND2019}, and on the other hand, there are also diagrams that differ significantly from those had been already encountered before.

\section*{Funding}

The work of P.\,E.\,Ryabov and S.\,V.\,Sokolov was sup\-por\-ted by RFBR grant 20-01-00399.

\section*{Conflict of interest}

The authors declare that they have no conflicts of interest.

\section*{Acknowledgements}

An explicit reduction to a system with one degree of freedom (the section 4) in the case of intensities of $\Gamma_1$ and $\Gamma_2$ of \textit{any} signs is made by G.\,P.~Palshin. The construction of a bifurcation diagram implicitly given by the relations \eqref{eq3_2} -- \eqref{eq3_31}, as well as the level lines of the reduced Hamiltonian \eqref{z3} was done with the help of program module on the \textit{Wolfram Mathematica} designed by G.\,P.\,Palshin.

\end{document}